\documentclass[nonacm, sigconf]{acmart}

\renewcommand\footnotetextcopyrightpermission[1]{} 
\usepackage{url}
\usepackage{breakurl}
\usepackage{hyperref}
\usepackage{graphicx}
\usepackage{balance}
\usepackage{booktabs}
\usepackage{subfigure}
\usepackage{algorithm}
\usepackage{algorithmicx}
\usepackage{algpseudocode}
\usepackage{setspace}
\usepackage{color}
\usepackage{enumitem}

\AtBeginDocument{%
  \providecommand\BibTeX{{%
    \normalfont B\kern-0.5em{\scshape i\kern-0.25em b}\kern-0.8em\TeX}}}

\newenvironment{myitem}{
\begin{itemize}[leftmargin=*]
 \setlength{\parskip}{0pt}
 \setlength{\itemsep}{1pt}
 \setlength{\partopsep}{0pt}
 \setlength{\parskip}{0pt}
 \setlength{\topsep}{0pt}
 \setlength{\parsep}{0pt}}{\end{itemize}
}

\begin{document}

\title{Contact Tracing: Beyond the Apps}

\author{Mohamed F Mokbel, Sofiane Abbar, Rade Stanojevic}
\affiliation{
  \institution{Qatar Computing Research institute}
  \institution{Hamad Bin Khalefa University, Doha, Qatar}
  \city{Email: \{mmokbel,sabbar,rstanojevic\}@hbku.edu.qa}}


\sloppy

\begin{abstract}
    
As pandemic wide spread results in locking down vital facilities, digital contact tracing is deemed as a key for re-opening. However, current efforts in digital contact tracing, running as mobile apps on users' smartphones, fall short in being effective. This paper lays out the vision and guidelines for the next era of digital contact tracing, where the contact tracing functionality is moved from being personal responsibility to be the responsibility of facilities that users visit daily. A privacy-preserving architecture is proposed, which can be mandated as a prerequisite for any facility to re-open during or after the pandemic.

\end{abstract}
\maketitle

\section{Introduction}
\label{sec:intro}

\noindent \textbf{Human-based Contact Tracing.} For the last few decades, human-based contact tracing has been vital in stopping the spread of infectious diseases~\cite{EK04}. Once a patient is confirmed positive, a community health worker talks to the patient to learn about other people who were in recent contact with the patient to screen them for the disease symptoms~\cite{WHO17}. Such simple human-based process had significant impact in saving lives by early diagnosing patients and avoiding further disease spread for tuberculosis (TB)~\cite{CDC13}, sexually-transmitted disease~\cite{Cla98}, Severe Acute Respiratory Syndrome (SARS)~\cite{LCC+03}, foot-and-mouth-disease~\cite{KGK05}, smallpox~\cite{PHF+04}, Ebola~\cite{WHO15}, among others. 

\noindent \textbf{The Need for Digital Contact Tracing.} With a pandemic wide spread and a worldwide lock down, causing unprecedented economic crisis, the main question that politicians, economists, and physicians are looking for is how and when we can re-open business while avoiding more pandemic wide spread. Since contact tracing is identified as a must for pandemic control~\cite{FWK+20}, the EU Commission recommends that contact tracing is needed for people to return to hotels and camping sites~\cite{CTPost20}. In USA, several states have made contact tracing a prerequisite for re-opening, including California~\cite{Cal20}, Pennsylvania~\cite{Sen20}, Virginia~\cite{WTOP20}, among others~\cite{NBC20}. Unfortunately, human-based contact tracing does not scale up to pandemic cases, with unknown immunity, high mortality rate, high reproduction number, and where contacts could be unknown to the patient, e.g., people met in airports, malls, or restaurants~\cite{SLX+20}.

\noindent \textbf{The Rise of Digital Contact Tracing.} Motivated by the limitations of human contact tracing, several governments have partnered with IT industry to come up with digital contact tracing techniques. The result is hundreds of mobile contact-tracing apps~\cite{OMJ20} where users would need to download the app and enable bluetooth connection and/or GPS location. Then the app uses either one or both of the following approaches:

\begin{myitem}

\item \textbf{Bluetooth User-to-user Contact Tracing.} The user is given a token ID to use every few hours. Once two users were in contact, their bluetooth connections will recognize each other ID and save it in a phone log. If a user is tested positive for a pandemic, authorities get access to the log and determine the IDs that were in close contact, and act accordingly. 

\item \textbf{GPS Location-based Contact Tracing.} Users periodically log their locations with the running app. Once a user is confirmed positive, authorities get access to the locations and find the recent user whereabouts. This information is propagated to all app users. Then, warning messages and authority notification will take place for users who were in the same location and time with the patient.

\end{myitem}

Though both approaches ensure user privacy through using pseudonym IDs~\cite{CFG+20}, the bluetooth approach ensures more privacy by not reporting user locations. Meanwhile, the bluetooth approach may miss some of the contacts of people who were in the same place with a confirmed patient, but only few minutes apart. On the other side, the GPS approach may end up reporting false positives in terms of higher number of potential cases that are not in risk. Yet, the GPS approach helps in identifying risky locations with higher virus spread.

\textbf{Our Vision.} In Section~\ref{sec:apps}, we make the case that app-based contact tracing does not work. Hence, in  Section~\ref{sec:guidelines}, we lay out our vision and guidelines for the next era of digital contact tracing. We believe that contact tracing should not be made as user responsibility and should not be running on user phones. Instead, contact tracing should be the responsibility of facilities and business entities (e.g., work places, malls, stadiums, restaurants, subways, etc) where the ability to do contact tracing can be made as a prerequisite for these facilities to re-open. In Section~\ref{sec:vision}, we outline the architecture that can realize our vision in a privacy-preserving way.

\section{Contact Tracing Apps: It does NOT work}
\label{sec:apps}

Unfortunately, even though there are tremendous efforts put in developing app-based contact tracing, it did not deliver what it has promised, mainly for the following two reasons: 

(1)~{\em Need for large cooperating population.} One of the very first apps, TraceTogether from Singapore~\cite{Trace}, has only reached around 1.4M users (25\% of population) after more than two months of release. This means that the probability that two random people in contact have both installed the app is only 6.25\%(0.25*0.25). This is assuming the best case scenario in which all users who have the app running in the background. With this tiny ratio, there is not much real benefit of such apps~\cite{TST20}. Meanwhile, though Iceland is reported to be the country with the highest population ratio using a contact tracing app (38\%), that did not help much~\cite{Business20}. (2)~{\em Low and biased smart phone penetration.} Smart phone penetration varies across countries, e.g., 24\% in India, 81\% in USA and 95\% in S. Korea~\cite{Stat20}. This leaves a major part of the population without access to app-based contact tracing~\cite{CCC20}. More importantly, smart phone penetration is inversely biased with COVID-19 spread. More poor areas has higher COVID-19 ratio~\cite{Bro20} and much less smartphone penetration, hence less access to contact-tracing apps.

So, unless contact tracing apps are made mandatory and used by the very large majority of population, they will not be effective~\cite{HBR20,Vox20}. With such serious issues, it becomes apparent that current efforts in digital contact tracing fail to meet the expectations. As a result, thoughts are going back to use human-based contact tracing, especially in USA, where it is estimated that 300,000 human contact tracers are needed for COVID-19~\cite{ASTHO20,CBS}.


\section{Next Era of Contact Tracing: Guidelines}
\label{sec:guidelines}

Our vision for the next era of digital contact tracing goes beyond mobile apps to be along the following guidelines:

\begin{myitem}

\item \textbf{Focus on unknown contacts.} 
Human tracers can efficiently identify family members, friends, or neighbors, but cannot identify unknown contacts who the patient have contacted in public facilities, e.g., malls, restaurants, work places. This should be the main focus of digital contact tracing.

\item \textbf{Focus on indoor environments.} People spend most of their time indoors, where it is more likely to get infected~\cite{BusinessInsider}. In USA, a national survey shows that people spend 87\% of their time indoor and 5\% in a vehicle~\cite{KNO+01}. Digital contact tracing should put more focus on indoor facilities. 

\item \textbf{Contact tracing is not a personal responsibility; it is surveillance.} People should not download any apps. Instead, each facility should be responsible on its own contact tracing for its visitors. If someone is tested positive, authorities will contact recently visited facilities and get their log of the patient contacts. The ability to do contact tracing would be a prerequisite for any facility to re-open. This is similar in principal that facilities will not re-open unless they comply with new hygiene guidelines. This also goes inline with the requirement that facilities should have enough CCTV camera coverage to ensure safe operations, where authorities will get access to, when accidents happen.

\item \textbf{Context-Aware tracing.} Digital contact tracing needs to go beyond the idea of one size fits all (same app running for everyone everywhere) to the more general case of context-awareness in terms of both infrastructure and analysis. For infrastructure, each facility may decide on its own way of deploying contact tracing technology. Meanwhile, the analysis of whether two persons were in contact would depend on the facility type. 

\item \textbf{Privacy-preserving.} Ensuring healthy environment should not be traded with privacy. Contact tracing should ensure that facilities do not have access to any user private information. 

\end{myitem}

\begin{figure}[t]
 \centering
  \includegraphics[width=0.5\textwidth]{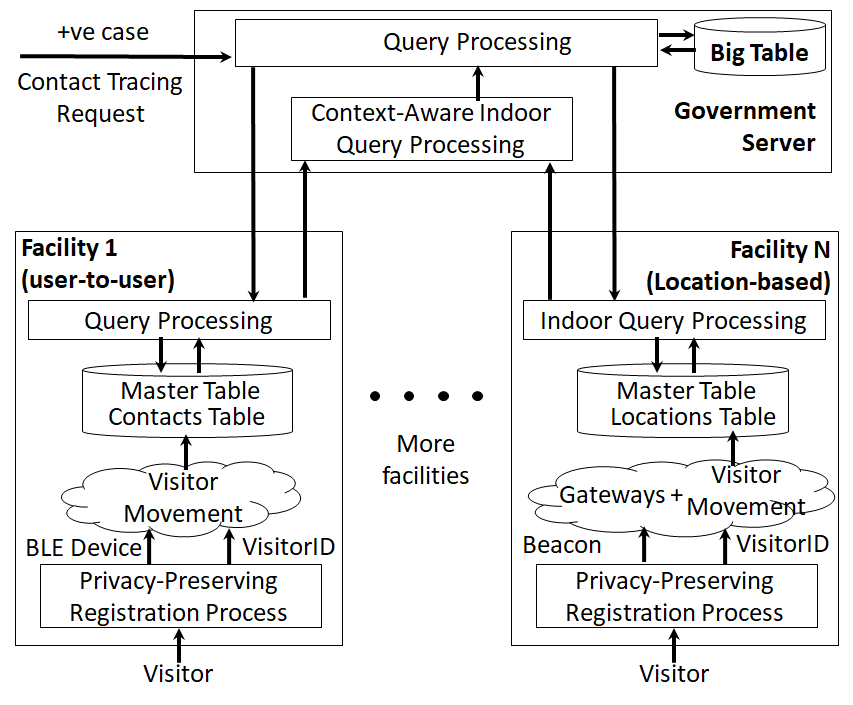}
  \vspace{-15pt}
 \caption{System Architecture. Facility 1 deploys {\em user-to-user} contact tracing, Facility N deploys {\em location-based} contact tracing. More facilities can be added independently.
 }
 \vspace{-5pt}
 \label{fig:arch}
 \end{figure}

\section{The Vision for the Next Era of Contact Tracing}
\label{sec:vision}

Figure~\ref{fig:arch} gives the system architecture of our vision for the next era of digital contact tracing. Each facility will decide on deploying one or both of the two approaches described in Section~\ref{sec:intro}, which we refer to as {\em User-to-User} and {\em Location-based Contact} Tracing. Then, each facility will have its own built-in infrastructure (Section~\ref{subsec:infra}), privacy-preserving registration process (Section~\ref{subsec:registeration}), and stored data structure (Section~\ref{subsec:datastructure}), which may be different based on the underlying infrastructure.

Once a person is identified positive with a pandemic, a contact tracing procedure is triggered at a government-owned server to identify the set of recently visited facilities (Section~\ref{subsec:procedure}). For each of these facilities, the server issues an API call that will trigger the facility query processing module to return a set of candidate contacts. An additional (optional) context-aware contact tracing analysis module (Section~\ref{subsec:context}) can be applied on the server side to get more accurate contact tracing information. 

\subsection{Infrastructure}
\label{subsec:infra}

The underlying infrastructure would be significantly different for {\em User-to-User} and {\em Location-based} contact tracing as follows:

\noindent \textbf{User-to-User Contact Tracing.} Upon entering a facility, visitors will be given a Bluetooth Low Energy (BLE) gateway in the form of a detached device, wristband, or key chain, that will be handed back before leaving. Examples of such devices are here~\cite{BLEGateway}. Each of these devices continuously broadcasts its own ID while reading the IDs of nearby devices. The reading log stored at each device would have the form {\em (BLE\_ID, timestamp, signal\_strength)} which presents the ID of the nearby device, along with the time and signal strength of reading that device. The latter is used as an indication of how close is the nearby device.

\noindent \textbf{Location-based Contact Tracing.} Upon entering a facility, visitors will be given a small Beacons device that would be returned before leaving. Examples of such devices are here~\cite{Beacons}. Unlike the BLE gateway devices in {\em user-to-user} contact tracing, Beacons devices: (a)~have much smaller size, and (b)~only broadcast their own IDs, but do not read any signal. Meanwhile, the facility will have several gateway readers fixed on the walls or ceiling that read broadcasted data from the Beacon devices in the form of {\em (BLE\_ID, timestamp, signal\_strength)}. Examples of such gateways are here~\cite{Gateways}.

\subsection{Privacy-Preserving Registration Process}
\label{subsec:registeration}

The privacy-preserving registration process is the same for both {\em user-to-user} and {\em location-based} contact tracing. A facility visitor will need to sign-in upon entry using a government-owned machine by entering her phone number or government ID. The machine will immediately generate a unique random ID that is given to the facility in exchange of the BLE or Beacon device. The random ID will be sent either as SMS to the visitor phone to ensure it is an actual number or as SMS to the machine itself in case the visitor has provided government ID instead of phone number. This means that the facility knows nothing about the visitor personal data. To the facility, the visitor is just a government-generated unique random ID. Furthermore, government and facilities will completely wipe any data is more than two weeks old.

Frequent visitors to a facility, e.g., employees at a work place, frequent airport travelers, or loyal store customers, may need to do the sign up process only once, where their phone numbers will be linked with their employee IDs or loyalty numbers. Then, the BLE or Beacon devices can be given to them once and actually attached to their work IDs or loyalty cards that they have to scan upon entering the facility.

\subsection{Stored Data Infrastructure}
\label{subsec:datastructure}

The data structure stored on the government-owned server is independent from the underlying infrastructure of each facility. It is basically one big table with the schema {\em (PhoneID, FacilityID, VisitorID, timestamp)}, which indicates that a user with a certain phone or ID number has visited a certain facility ID at a certain time, and was given a certain visitor ID. For efficient retrieval, the table is accessed through two hash tables for PhoneID and VisitorID.

Meanwhile, each facility, regardless of the underlying contact tracing infrastructure, maintains a {\em Master} table with the schema {\em (VisitorID, BLE\_ID, time\_in, time\_out)}, which indicates that a certain BLE or Beacon ID was given to a certain visitor within a certain time frame. The {\em Master} table is accessed through two hash tables for VisitorID and BLE ID. In addition, each facility maintains the following data structure(s), based on the underlying infrastructure:

\noindent \textbf{User-to-User Contact Tracing.} Each facility maintains a {\em Contacts} table: {\em (BLE\_ID1, BLE\_ID2, timestamp, signal\_strength)}, which logs the timestamp and signal strength for each pair of BLE devices that came close to each other. The signal strength is converted to some universal distance measure (e.g., meters) to accommodate that different devices may present signal strength differently. The table is populated by combining all readings received from individual BLE devices, and is accessed through a hash table over BLE\_ID1.

\noindent \textbf{Location-based Contact Tracing.} Each facility maintains a {\em Locations} table with the schema {\em (BLE\_ID, location, timestamp)}, which logs the locations of each BLE within the facility at a certain timestamp. The indoor location is not a traditional {\em <lat,log>} coordinates. Instead, it is more of a symbolic descriptive location based on the facility map~\cite{Sam19}. The {\em Locations} table accessed by a hash table over BLE\_ID, and is populated through a typical trilateration process where the readings from three fixed Gateways for the same BLE is used to come up with the symbolic BLE location at a certain time~\cite{IoTIndoor}. Such process has been commonly used in indoor positioning systems for real-time asset tracking~\cite{TrackingIndoor}.

\subsection{Contact Tracing Procedure}
\label{subsec:procedure}

Once a person is confirmed positive for the pandemic, the contact tracing procedure is triggered on the server side by government officials. A simple local query with the patient phone (or ID) and a certain time period (e.g., last two weeks) would return the set of facilities $F$ visited by the patient, with the anonymized visitor ID and timestamp of each visit. A patient may have visited the same facility multiple times, each with a different visitor ID. For each facility in $F$, the server sends an API call inquiry asking for all visitor IDs who were in contact with the patient visitor ID.

Whenever a facility receives a query with a visitor ID and a timestamp, it uses its {\em Master} table to map the visitor ID to the BLE ID used during the visit, along with the visit time frame. Then, based on the underlying infrastructure, the following information is retrieved and sent back to the server: 

\noindent \textbf{User-to-User Contact Tracing.} Given a BLE ID and visit time frame, the facility will use its {\em Contacts} table to retrieve all other BLE IDs that were reported in contact with the visitor BLE ID, along with the timestamp and estimated distance of each contact event. Then, a reverse lookup over the {\em Master} table will get the corresponding Visitor ID for each contacted BLE ID. The information sent back to the requesting authority server will have the schema {\em (VisitorID, timestamp, estimated\_distance)}.

\noindent \textbf{Location-based Contact Tracing.} Given a BLE ID and visit time frame, the facility will use its {\em Locations} table to retrieve the trajectory of locations (with timestamps) within the facility during the visit. Then, a spatio-temporal indoor range query~\cite{LCJ12} would retrieve all the BLE IDs that were in a close spatio-temporal proximity to the given BLE ID. The parameters of spatio-temporal proximity are set in a conservative way, e.g., within 10 meters distance and 10 minutes time frame. Then, a reverse lookup over the {\em Master} table will get the corresponding Visitor ID for each nearby BLE ID. Finally, the information sent back to the requesting authority server is: {\em (VisitorID, location, timestamp, spatial proximity, temporal proximity)}. In addition, each facility may optionally send the full spatio-temporal trajectory of the patient visitor within the facility, which can be used for further analysis at the requesting authority.

\subsection{Context-Aware Contact Tracing Analysis}
\label{subsec:context}

When the government-owned server receives back the results from each facility, it may just use the list of contacts or nearby visitors as the ones in risk. In this case, a reverse lookup with the Visitor ID over the local server table would return the phone number (or ID) of each visitor. Health officials can take it from there and start contacting the people accordingly. However, as a means of increasing the accuracy, additional context-aware contact tracing analysis can be employed based on the underlying infrastructure: 

\noindent \textbf{User-to-User Contact Tracing.} The signal strength of each contact may be interpreted differently based on the facility type. For example, within a stadium, one may focus on the readings with high signals. Within a restaurant, one may even report lower signals, only if they were persistent over a certain period of time. Within a Mall, a different search criteria and parameters can be used.

\noindent \textbf{Location-based Contact Tracing.} Assuming the availability of facility layout, the spatial and temporal proximity of visitors may be interpreted differently based on the facility type and layout. Two contacts who are close by spatially and temporally may have a wall in between, and hence the proximity is not risky. Meanwhile, a heatmap may be depicted for the facility indicating regions of high risk, where contact information may be interpreted differently. Furthermore, depending on the nature of the pandemic and how it spreads (e.g., via surface or air), we can find users who have been to the spots recently visited by a patient. For example, a patient who uses a table in a food court, leaves it, then another visitor uses the same table.

Generally speaking, there are many context-aware indoor analysis that can be deployed~\cite{ARI+12}, though there are way more rich analysis for the case of {\em location-based} contact tracing than {\em user-to-user} contact tracing. Having the context-aware analysis module on the server side instead of having it on each facility is mainly to allow health officials to change the parameter settings and search criteria without the need to get back to the facility. Another alternative is to have such analysis on the facility side, accessed via more sophisticated API calls that account for more parameters such as minimum distance, contact time interval, and location label.

\section{Conclusion}
\label{sec:Conclusion}

The paper makes the case that current app-based contact tracing techniques are not effective. Then, the paper lays out the vision for the next era of digital contact tracing where the responsibility of contact tracing is moved from the persons to the facilities that the persons visit. Each facility, e.g., mall, work place, stadium, train, restaurant, should have the ability to do contact tracing for all its visitors. Such ability could be enforced as a prerequisite for any facility to re-open during a pandemic. A privacy-preserving architecture and infrastructure that achieve such vision is presented. The architecture allows each facility to independently decide whether to deploy a {\em user-to-user} or {\em location-based} contact tracing approach. The former approach mainly reports the people in contact to the patient, while the second approach additionally reports the locations of the patient and the contacts.

\balance

\bibliographystyle{abbrv} 
\begin{scriptsize}
\bibliography{M_ContactTracing}
\end{scriptsize}

\end{document}